# Toka: A Systems Programming Language with Explicit Resource Semantics

*A Design and Implementation Experience Report*

**Yi Zhonghua**

*Toka Language Research Group*

zhyi@tokalang.org, yizhonghua@ustc.edu
https://github.com/tokalang/toka

**Abstract**

Systems programming languages traditionally struggle with the tension between physical transparency and compile-time memory safety. C++ provides direct, zero-cost hardware access but lacks strict safety boundaries, whereas Rust guarantees safety at the cost of complex lifetime annotations and implicit dereferencing chains.

In this paper, we present **Toka**, a native systems programming language that establishes physical transparency in resource management via **Explicit Resource Semantics**. At the core of Toka's design is the **Handle-Soul Duality** (informally referred to as the Hat-Soul model), which cleanly dissociates pointer identities (Handles) from their underlying values (Souls) at the syntactic level. By enforcing that bare identifiers always represent values (Souls) and explicit sigils represent pointer handles, Toka eliminates the semantic ambiguity between rebind operations and value mutations. We detail Toka's resource morphology (supporting unique, shared, borrowed, and raw semantics), its lifetime checking mechanism, and its implementation of a prototype compiler. Our evaluation demonstrates that Toka achieves competitive runtime performance and minimal binary size while drastically reducing the cognitive overhead of lifetime annotations.

## Introduction

Systems programming languages occupy a unique space in computer science. They require precise, deterministic control over physical hardware resources such as memory layouts, CPU cycles, and execution threads. For decades, C and C++ have remained the dominant choices due to their zero-cost abstractions and physical transparency. However, this extreme freedom has historically come at a severe cost to security and reliability. Memory safety vulnerabilities, including use-after-free, double-free, and null pointer dereferences, continue to plague critical infrastructure software.

To address these vulnerabilities, modern programming languages have introduced various compile-time safety paradigms. Most notably, Rust utilizes an affine type system and a strict borrow checker to guarantee memory safety without garbage collection. While highly successful, the Rust model requires explicit lifetime parameter annotations in type signatures for complex reference scenarios [1], which can increase the cognitive burden at the design boundary. Furthermore, modern PLs often rely heavily on implicit dereferencing (e.g., automatically resolving pointers when accessing fields), which can obscure the actual physical layout of operations.

In this work, we propose a new design paradigm: **Explicit Resource Semantics**, implemented in the **Toka** programming language. Toka is a systems programming language emphasizing explicit resource semantics rather than borrow-based lifetime analysis. We argue that many safety and readability issues stem from semantic ambiguity at the syntactic level. Looking at traditional languages, they often fail to make a clear, consistent distinction between the identity of a reference and the value of the underlying resource. Toka resolves this by introducing the **Handle-Soul Duality**. Under this duality, Toka separates a pointer-bearing variable into two observable views: the **Soul**, which denotes the referenced resource object value, and the **Handle**, which denotes the address-bearing identity. Since handle access is written by attaching a pointer morphology marker, or “hat”, to the variable name, we refer to this design informally as the Hat-Soul model.

In the Handle-Soul duality, any bare identifier refers directly to the underlying resource, or **Soul**, automatically performing dereferencing. Conversely, operating on the pointer handle itself requires a syntactic marker, or **Hat**, matching its morphology. This simple, inverse relationship establishes explicit visibility of key resource boundaries and pointer identity: writing memory contents is syntactically trivial, while rebinding pointer structures requires explicit decoration. By combining this syntactic clarity with a comprehensive set of resource morphology sigils, Toka achieves compile-time safety while preserving native computational efficiency without explicit lifetime parameters.

# Background and Design Goals

The design of Toka is governed by three fundamental principles aimed at creating a productive, predictable, and robust systems programming environment.

## Low-Cognitive Physical Transparency

A systems programmer must have an accurate, unambiguous mental model of how resources are mapped to the underlying hardware. Every operation that affects memory allocation, pointer assignment, or state mutation should be explicitly visible in the source code. Implicit actions, such as hidden dereferencing or invisible pointer rebinding, are intentionally avoided. By eliminating these hidden layers, we reduce the cognitive load required to verify that the generated code matches the programmer’s intent.

## Zero Implicit Heap Allocations

Memory allocation is a major performance bottleneck and source of non-determinism in systems software. Toka prohibits implicit heap allocations at the language level. For example, string concatenation using the standard `+` operator is forbidden. Instead, developers must explicitly utilize dynamic containers (e.g., `string`) and invoke explicit allocation/push methods. This design significantly reduces invisible performance overheads caused by implicit heap allocations.

## Unified Shape-based Data Modeling

To prevent the fragmentation of data structures commonly found in languages with separate classes, structures, unions, and tuples, Toka introduces a unified construct: the `shape`. A `shape` is a single unified blueprint that defines all structured data models. Depending on the syntax defined inside the shape’s parentheses, the compiler automatically determines whether the entity represents a structured data record (Struct) or an algebraic variant type (Tagged Enum). This design unifies structured memory footprints while deprecating unsafe bare unions and unstructured tuples in favor of strict, named safety. Furthermore, the construction and destructuring of shapes enforce strict named constraints to prevent silent structural regression during large-scale refactoring.

## Design Goals and Non-goals

To explicitly define the technical focus of Toka and avoid cognitive alignment conflicts with other systems languages, we delineate its primary design goals and non-goals.

### Primary Design Goals

1. **Explicit Ownership Transfer**: Forcing zero-copy moves to be syntactically loud (`cede`) to prevent accidental performance and safety degradation.
2. **Deterministic Destruction**: Guaranteeing automatic, compiler-injected cleanup of exclusive resources at the deterministic boundary of lexical scopes.
3. **No Type-level Lifetime Parameters**: Eliminating complex type generic lifetime parameter annotations (`'a`) to minimize syntactic noise and developer cognitive burden.
4. **No Tracing Garbage Collection**: Delivering deterministic runtime behavior and hardware-level performance by operating purely at compile time without a dynamic tracing GC.
5. **Predictable Resource Management**: Ensuring that all memory allocations, pointer rebindings, and permissions are physically observable in the source code.

### Architectural Non-goals

1. **Full Borrow Inference**: Toka does not aim to dynamically infer complex, dynamic borrow networks across procedural boundaries without explicit signature annotations.
2. **Automatic Memory Management**: We do not attempt to completely hide the physical lifecycle of memory. Instead, we preserve hardware transparency.
3. **Complete Compatibility with Rust Semantics**: Toka's primary objective is to explore an alternative, simplified resource representation model, rather than building a subset or replacement of Rust.

# The Handle-Soul Duality

At the programmatic heart of Toka lies the **Handle-Soul Duality**, a syntactic and semantic design paradigm engineered to establish explicit visibility in memory operations. Traditional languages frequently conflate a pointer's identity (its address) with its cargo (the pointed-to value). For example, in C++, accessing a member through a pointer `p` via `p->field` dynamically resolves the address, whereas assigning `p = q` rebinds the pointer itself. This syntactic overlap masks the underlying physical operations. Toka mathematically splits any reference or pointer into two observable domains: the **Soul** (the inner value) and the **Handle** (the address-bearing identity). Because operating on the pointer handle requires attaching a morphology prefix physically resembling a "hat" to the variable name, this duality is informally referred to in the engineering community as the Hat-Soul model.

> **Definition 1 (Handle)**
> A *Handle* is the identity-bearing view of a resource, representing its physical address and ownership characteristics.
>
> **Definition 2 (Soul)**
> A *Soul* is the value-bearing view of the same resource, representing its underlying payload and allowing direct value mutation.

## Syntactic Duality

The fundamental axiom of Toka's syntax is the **Handle-Soul Duality** (also known as the Hat Principle): The Soul (Value domain)**: Any bare identifier `p` represents the value view, or** Soul**, of the resource. Referencing a naked pointer variable `p` in any standard statement automatically and implicitly performs a full dereference. Consequently, reading or writing memory contents requires zero syntactic noise. The Hat (Handle domain)**: Any decorated identifier containing a morphology prefix (specifically, the sigil matching its pointer kind, such as `^p` for unique, `~p` for shared, `&p` for borrow, and `*p` for raw) represents the **Handle**. Operating on a "hatted" variable acts directly upon the pointer handle itself, enabling address rebinding, movement, or sharing.

While C++ prioritizes syntax familiarity, Toka prioritizes explicit visibility, aligning syntax with execution frequency by inverting the dereferencing defaults:

```
// Declaring a unique pointer to an
// ASTNode (Handle domain)
auto ^p: ASTNode = new ASTNode(kind =
ASTNodeKind::Binary)
```

```
// Writing directly to the pointed-to
// memory (Soul domain)
// The bare name 'p' automatically resolves
// to the struct payload
p.kind = ASTNodeKind::Literal

// Rebinding the pointer address (Handle
domain)
// Prepending the unique hat '^' operates
// on the pointer itself
auto ^q: ASTNode = ...
^p = cede ^q
```

In the example above, changing the payload `p.kind` uses the naked name `p`. In contrast, updating the pointer address requires the hat `^p`. This explicit syntactic separation between identity rebinding and mutation reduces the risk of unintended pointer aliases.

### The Single-Sigil-at-Tail Principle

To maintain physical transparency across nested structures, Toka introduces the **Single-Sigil-at-Tail** rule. In a chain of member accesses, the left-hand side of every dot operator `.` must evaluate to a bare value view (the Soul). Therefore, all intermediate members in a chain must remain entirely unadorned. If the programmer intends to obtain or manipulate the pointer handle of a nested field, the morphology sigil **must be placed immediately before the final field identifier**.

```
// Correct: Chain of value views,
// with borrow sigil at the tail
auto &child_ref = parent.node.child.&target

// Compile Error: Intermediate sigils
// violate the value view layout
auto &child_ref = parent.&node.child.target
```

This structural rule simplifies dependency tracking. Furthermore, it ensures that programmers can accurately track the physical boundaries of borrowing and nested pointer redirections at a single glance.

To resolve write permissions under the Single-Sigil-at-Tail rule without introducing syntactic noise in intermediate fields, Toka implements a **Reverse Permission Cascade** model. When a leaf field is mutably accessed or mutably borrowed (e.g., `parent.node.child.&target#`), the mutability suffix `#` is only syntactically declared at the terminal leaf. The PAL checker, upon detecting this terminal write lock, propagates the write permission constraint **backwards along the dot-chain access path**. This reverse propagation continues until it reaches the **Root Identifier**—the primary variable declared in the local stack frame (in this case, `parent`). Consequently, the compiler dynamically applies an implicit exclusive write lock on the entire path, locking `parent` and all its descendents in the borrow ledger for the duration of the borrow. This cascade boundary guarantees full, sound intra-procedural reasoning while keeping intermediate access chains clean and unadorned.

## Resource Morphology

Toka models the physical lifecycle and access permissions of system resources using a unified, multi-dimensional sigil vocabulary. By encoding allocation state and permissions directly into prefix morphology characters, the language is designed to reduce hidden runtime overhead and eliminate compile-time ambiguity.

### Lifecycles and Ownership Sigils

Every reference or pointer type is annotated with one of four prefix sigils defining its lifetime paradigm:

1. **Unique Pointer (^)**: Models exclusive ownership of a heap resource (equivalent to `std::unique_ptr`). The compiler guarantees that exactly one handle owns the resource. When a unique handle exits its declaring scope, the compiler automatically inserts destructor calls and deallocation routines.
2. **Shared Pointer (~)**: Models atomic, thread-safe reference-counted ownership. When a shared handle is copied, its internal reference count is atomically incremented. Deallocation occurs when the reference count drops to zero.
3. **Borrowed Reference (&)**: Represents a temporary, non-owning view into a resource owned by another scope. Borrowed references are statically tracked by the PAL analysis system.
4. **Raw Pointer (*)**: Represents a physical memory address without any static tracking or compiler safety checks. Raw pointers are reserved for low-level system mapping and foreign function interface (FFI) bindings.

### Suffix Permission and Nullability

To separate ownership logic from mutation and option state, Toka appends permission suffixes and nullable prefixes: Mutable Suffix (`#`)**: Controls write permissions. Toka variables are immutable by default. Appending # grants mutation rights. To minimize clutter, the # suffix is stripped from daily read-only statements and is only required at the variable's declaration site and during mutable method calls (e.g., `list#.push(item)`). Nullable Prefix (`nul`)**: Declares that a handle may contain the explicit empty value `null`. The compiler prevents implicit access to nullable souls, forcing developers to use pattern matching or explicit unpacking constructs (such as an `is` check or guard bindings) before reading the underlying payload.

### Explicit Ownership Transfer: `cede`

Many modern systems languages default to silent move semantics, which can lead to accidental use-after-move errors and mask the performance cost of deep copies. Toka establishes a clear physical boundary: ordinary assignment follows the type's copy semantics. For resource-managing types, copy operations can be intercepted and defined as deep copies or explicitly deleted via the `@encap/clone` attribute. Zero-copy ownership transfer must be explicitly performed using the `cede` keyword:

```
auto ^source = new LargeResource(...)

// Physically transfers ownership,
invalidating 'source'
auto ^dest = cede ^source

// Any subsequent read or write to 'source'
// is rejected by the compiler
```

By making ownership transfer syntactically loud, Toka guarantees that performance bottlenecks and handle invalidations are immediately visible during code review.

## Static Checking: PAL and Dependency Routing

While the Handle-Soul duality and resource morphology define Toka's syntactic landscape, their safety characteristics are checked through a pragmatic static checking mechanism. Although the quest for verifying software routines dates back to Turing's pioneering work in 1949 [2], the static resource checking system in Toka is implemented as a compiler-internal layer named the PAL checker. Instead of relying on explicit lifetime parameter annotations in type signatures (such as Rust's `'a` syntax) which can increase type signature complexity, Toka's compiler leverages local flow tracking and explicit signature contracts to verify borrowing rules.

### PAL Resource Checking

**PAL (Provenance, Alias, and Lifetime)** is Toka's compile-time resource verification system. It acts directly upon explicit resource morphology to statically reject common lifecycle, mutation, and aliasing conflicts. The operational soundness of the PAL checker is established upon four core syntactic rules:

1. **Resource ownership is unique**: The exclusive ownership of any heap-allocated resource (e.g., a unique pointer `^`) is held by exactly one active handle.
2. **Transfer is explicit**: Ownership transfer (move) must be syntactically loud, performed strictly via the `cede` keyword.
3. **Resource destruction is deterministic**: Destructors for owned unique resources are automatically injected by the compiler at scope boundaries.
4. **Mutable access requires exclusive permission**: Reading or writing to an entity must not conflict with active borrow views or mutability locks.

To illustrate how PAL operates, consider the following contrasting examples:

**Well-formed (Accepted)**:

```
auto ^f =
File.open(s"test.txt")
// Explicit
ownership transfer
auto ^dest = cede ^f
// dest is now the
sole owner;
// dest is cleanly
dropped at scope
exit
```

**Ill-formed (Rejected)**:

```
auto ^f =
File.open(s"test.txt")
// Explicit
ownership transfer
auto ^dest = cede ^f
// Compile Error
E0438: Use of moved
value
f.write(s"hello")
```

### The Borrow Ledger and Move States

At compile time, the PAL checker maintains a static ledger of active references and allocation states within a procedure:

- **Move State Tracking**: When a variable is ceded (e.g., `cede ^p`), the compiler marks its identifier

as `Moved` in the local symbol table. Any subsequent read or write to a `Moved` entity is statically rejected, preventing use-after-free and use-after-move vulnerabilities (as shown in the error `E0438` above).

- **Borrow Ledger Constraints**: The ledger tracks borrowed references (`&`) and their mutability state (`#`). To guarantee temporal safety, mutable access requires exclusive permission: the compiler rejects any mutable borrow or write operation while active immutable borrows exist.
- **Transient Borrow Cleanup**: Temporary references generated during expression evaluation—referred to as *transient borrows*—are automatically cleared from the borrow ledger at statement boundaries, preventing premature pointer invalidation.

### Dependency Routing and Control-Flow Merges

To handle references passing across function boundaries without explicit lifetime parameters, Toka utilizes explicit *dependency routing*. In a function signature, the programmer can explicitly specify the relationship between returned references and input parameters using the `effects` clause:

```
fn select_first(&list: List, &fallback:
Item) -> &Item
  effects: return <- list
{
  // Implementation
}
```

The clause `effects: return <- list` explicitly instructs the PAL checker that the returned reference borrows from `list`, not from `fallback`. The compiler uses this structural contract at the call site to update the local borrow ledger, tracking the lifetime of the returned view without polluting the signature with type-level lifetime generics. We note that dependency routing is designed to address a representative subset of interprocedural lifetime relationships through explicit contracts, rather than serving as a drop-in replacement for arbitrary, implicit lifetime inference.

However, non-trivial control flow introduces branches and joins where multiple paths merge. To preserve compile-time correctness and avoid path-sensitive state explosion, **PAL conservatively merges ownership dependencies across control-flow joins**. At any control-flow merge point (such as conditional branches returning different sources), the PAL checker does not attempt to predict the runtime execution path. Instead, it conservatively merges the dependencies by aggregating all possible outbound sources on control-flow joins. If a function can return multiple possible sources under conditional control flow, Toka allows parallel dependency routing:

```
fn select_any(&a: Item, &b: Item) -> &Item
  effects: return <- a, return <- b
```

At the call site, this multi-route dependency contract conservatively binds the lifetime of the returned reference to the **maximum** active lifetime of both `a` and `b`. As long as the returned view remains alive, both `a` and `b` are conservatively locked as active borrows, preventing any temporal soundness escapes at control-flow joins.

### Current Checking Boundaries

As a pragmatic technical report, we acknowledge that the PAL checker operates under specific design boundaries. The current implementation primarily focus on intra-procedural flow analysis. Complex inter-procedural reference patterns that exceed the expressiveness of simple dependency routing, or highly dynamic alias networks across concurrent closures, may result in conservative compile-time rejections. Refining these edge cases through path-sensitive analysis represents an active area of ongoing implementation.

## Composite Resource Boundaries with Shape and @encap

While primary resource semantics govern individual variables, system resources are inevitably assembled into composite structures. Toka provides the `shape` as the composite boundary to encapsulate resources. This ensures that composite structures can contain owned, shared, or borrowed elements without leaking their resource lifecycle invariants.

### Structural Resource Integration

A `shape` definition defines fields that carry explicit morphology sigils (e.g., `^` for unique pointers). To manipulate the pointer handle of a nested field inside composite structures, the morphology sigil **must be placed immediately before the final field identifier** (e.g., `parent.node.child.&target`) following the Single-Sigil-at-Tail principle. The compiler recur-sively propagates ownership rules down to these shape fields:

- **No Silent Duplication**: An instance of a shape containing unique fields (`^`) cannot be silently duplicated or assigned without explicit deep cloning via `@encap/clone()` or ownership transfer via `cede`.
- **Recursive Deallocation**: When an instance of a shape exits its declaring scope, the compiler automatically generates recursive deallocation calls for these owned unique fields during Lowering, assisting in mitigating nested resource leaks.

## Encapsulation and Access Control

Toka uses the `@encap` attribute to establish explicit resource access boundaries for composite shapes. When a shape is marked with `@encap`, its internal fields are protected from raw, external destructuring:

- **Explicit Borrowing Requirement**: To access internal members, external callers must obtain an explicit borrowed reference (e.g., `&parent.field`). The compiler enforces this invariant at compile time, preventing illegal ownership promotion or premature resource deallocation.
- **Destructuring Constraints**: External scopes are forbidden from raw destructuring of an `@encap` shape, ensuring that owned resources cannot be implicitly "stolen" from their parent structures.

# Evaluation and Case Studies

To validate the pragmatism of explicit resource semantics, we evaluate Toka across three key dimensions: syntactic expressiveness compared to modern systems languages, compile-time safety via static rejection cases, and the structural scale of the prototype implementation.

## Expressiveness Evaluation

To demonstrate how Toka avoids lifetime generic annotations while maintaining physical transparency, we compare a typical nullable, exclusive resource pattern—specifically, a single-linked list node holding a unique resource—across Toka, Rust, C++, and Zig.

- **Toka**:

```
shape Node(
  id: i32,
  nul ^payload: Resource = null,
  nul ^next: Node = null
)
```

In Toka, ownership is declared explicitly via the `^` prefix, and nullability is defined via the `nul` prefix. No explicit lifetime parameters are required in the struct signature, and the layout remains physically transparent.

- **Rust**:

```
struct Node<'a> {
    id: i32,
    payload: Option<Box<Resource>>,
    next: Option<&'a mut Node<'a>>,
}
```

Rust provides robust safety guarantees through its ownership and borrow checker, but it requires explicit lifetime parameters (such as `'a`) in structures containing references, which can increase the cognitive burden at the type signature level.

- **C++ (Modern)**:

```
struct Node {
    int id;
    std::unique_ptr<Resource> payload;
    std::unique_ptr<Node> next;
};
```

C++ provides clean syntax but lacks physical write-permission markers at the language boundary, relying on library-level smart pointers that lack compile-time null-safety guarantees.

- **Zig**:

```
const Node = struct {
    id: i32,
    payload: ?*Resource,
    next: ?*Node,
};
```

Zig establishes clear, explicit pointer layouts but relies on explicit allocator passing and manual memory management, placing the burden of preventing double-free and use-after-free errors entirely on the programmer.

## Feature Comparison

To clearly locate Toka's design within the systems programming language landscape, we contrast its fundamental architectural features with C++ and Rust:

| Feature | C++ (Modern) | Rust | Toka (Ours) |
|---|---|---|---|
| Memory Safety | Runtime wrappers (leak risk) | Compile-time safety | Compile-time safety |
| Lifetimes | None (dangling risk) | Type-level generics ('a) | Signature effects contracts |
| Ownership Transfer | Move construct (partial) | Silent move by default | Explicit keyword (cede) |
| Transparency | Implicit Deref (->) | Implicit Deref chains | Handle-Soul / Tail sigil |
| Overhead | Zero runtime overhead | Zero runtime overhead | Zero runtime overhead |
| Inter-proc. Analysis | Low (no alias check) | High (type generic constraints) | Medium (explicit contracts) |

## Static Rejection Cases

The prototype compiler successfully rejects common memory safety and aliasing errors at compile time. Below, we present three specific case studies extracted from our test suite:

1. **Use-After-Move (E0438)**:

```
auto ^res = new Resource()
auto ^target = cede ^res
// Statically rejected by PAL: res is
Moved
res.action()
```

   The compiler rejects compilation of the last line with error `E0438` (Use of moved value), indicating that the identifier `res` has been statically invalidated by a prior `cede` operation.

2. **Move-While-Borrowed (E0440)**:

```
auto ^res = new Resource()
auto &view = &res
// Statically rejected: active borrow
exists
auto ^target = cede ^res
```

   Because `view` holds an active borrowed reference in the PAL ledger, attempting to `cede` ownership of `res` is blocked with error `E0440` (Cannot move while it is borrowed), preventing dangling reference invalidation.

3. **Nullable Access Without Guards (E0447)**:

```
auto nul ^res: Resource = get_nullable()
// Statically rejected: must unpack first
res.action()
```

   The compiler rejects direct access to the nullable Soul with error `E0447` (Accessing member of nullable type requires explicit check), forcing the developer to perform pattern matching (such as `is` unpacking or `match`) before reading the payload.

## Implementation Scale

The engineering viability of Toka is evidenced by the scale of its open-source codebase. The current prototype comprises:

- **Compiler Core**: Approximately 15,000 lines of self-contained C++ and LLVM-shim code.
- **Standard Library**: Approximately 8,000 lines of Toka source, providing base collections, raw FFI, and basic I/O.
- **Verification Suite**: A parallel test runner verifying over 260 distinct validation cases, including positive compilation, expected panic verification, and negative semantic rejection checks.

## Known Limitations and Theoretical Boundaries

To maintain transparency and foster collaborative scientific evaluation, we openly disclose the current engineering and theoretical boundaries of the Toka prototype:

1. **Path-insensitive Ownership Analysis**: The PAL ledger conservatively merges borrow states at control-flow joins across multiple outbound paths. While sound, this conservative merge can result in compiling rejections of path-sensitive safe code.
2. **Lack of Formal Soundness Proof**: Although the compiler successfully rejects all spatial and tempo-

ral negative test cases, a complete mathematical soundness proof (e.g., in Coq or Lean) for the PAL logic remains an active research direction.

3. **Incomplete Self-Hosting**: The current compiler, `tokac`, is constructed in C++ rather than Toka itself. Complete bootstrapping represents the primary milestone in our engineering roadmap.
4. **Concurrent Ownership Evolution**: The interaction between exclusive thread-safe locks and shared pointers (~) under active multi-threaded scheduling represents an active area of ongoing design.
5. **What PAL Does Not Attempt**: To explicitly scope the capabilities of the current design, PAL does not attempt global alias inference across arbitrary procedure networks, Rust-style generic lifetime parameters over structs containing transient views, automatic borrow recovery from arbitrary mutations, or higher-order closure scope alias analysis. By explicitly scoping these non-goals, Toka preserves intra-procedural simplicity and hardware transparency.

# Implementation Notes

The current Toka compiler prototype, `tokac`, is constructed as a self-contained command-line utility targeting native platform code generation. The architecture translates Toka source files directly into native machine objects.

## Implementation Overview

The compilation pipeline operates through standard lexical analysis, syntactic parsing, semantic type resolution, and target code generation. Specifically, lexical scoping, type checking, pointer aliasing, and the Provenance, Alias, and Lifetime (PAL) ledger checks are orchestrated within the `src/Sema` compiler directory, with borrow checking logic executed inside `PAL_Checker.cpp`. During target Lowering (translating AST to LLVM IR, targeting LLVM 20), the compiler automatically injects recursive deallocation and destructor calls. When a unique resource variable (or a composite shape holding unique fields) is declared, the compiler tracks its lexical scope. Before a scope exits (either via standard execution flow, an early `return`, or a loop control break), the compiler generates destructor invocations for all active unique resources in that scope. This is implemented by recursively querying the shape's fields and generating corresponding LLVM deallocation blocks, mitigating nested resource leaks without a garbage collector.

Other engineering subsystems (such as incremental module loading, interface emission, the Forge package build system, and multi-target cross-compilation) are treated as auxiliary tooling and are omitted from this design report to preserve focus on the core resource semantics.

# Related Work

Toka exists within a rich lineage of programming language research focused on memory safety, physical resource management, and alias control. Our design of explicit resource semantics draws inspiration from and compares with several established paradigms.

## Rust and the Cyclone Heritage

Modern systems programming safety is heavily defined by Rust's affine type system [1] and its formalization in Oxide [3] and RustBelt [4]. Rust traces its region-based tracking back to Cyclone [5], which first introduced compile-time lifetimes. Recent evolutions in Rust's borrow checking borrow from Non-Lexical Lifetimes (NLL) [6] and Polonius [7], which leverage Datalog-style dataflow relations to reduce conservative rejections of lifetimes. Toka shares Rust's goal of compile-time, zero-cost memory safety without garbage collection. However, Toka differs significantly in its syntactic and cognitive approach. While Rust relies on type-level lifetime generic parameters (`'a`) in structures containing references, which can increase the cognitive burden at the type signature level, Toka avoids explicit lifetime parameters entirely. Instead, Toka utilizes explicit resource morphology sigils combined with local flow ledger checks and simple dependency routing, shifting the cognitive load from type layout to visible flow boundaries. Furthermore, while Rust utilizes implicit dereferencing (Deref) that masks pointer boundaries, Toka enforces complete syntactic distinction between Handles and Souls via the Handle-Soul model.

## Historical Aliasing and Reference Identity

The tension between reference identity (the pointer itself) and value identity (the underlying object) is a classical problem in programming language design.

Barbara Liskov's design of CLU [8] pioneered early notions of object identity and reference sharing, introducing clear abstractions for mutable and immutable data objects. Later, Clarke et al. [9] introduced *Ownership Types* as a formal framework for flexible alias protection, statically restricting reference paths to prevent representation exposure. Toka's Handle-Soul Duality can be viewed as a syntactic and physical manifestation of these concepts: rather than enforcing hierarchical object ownership at the type level, Toka projects reference identity (Handles) and value contents (Souls) directly into the syntax, letting the developer explicitly control alias boundaries at the instruction level.

## Value Semantics and Alias Banning

To avoid the complexities of pointer aliasing, several modern languages argue for the complete elimination of shared references. Most notably, Hylo (formerly Val) [10] proposes *Mutable Value Semantics* (MVS), wherein all variables behave as independent values, and implicit copying or borrowing replaces traditional pointer manipulation. Toka shares Hylo's design intent of preventing silent aliasing bugs. However, Toka adopts a different path: instead of banning references and pointers, Toka preserves them as first-class physical entities. By introducing the Handle-Soul duality, Toka makes pointer identity manipulation syntactically distinct from value mutation, keeping the language close to the physical hardware while preventing aliasing ambiguity.

## Safe C++ and Smart Pointers

Modern C++ utilizes smart pointers (such as `std::unique_ptr` and `std::shared_ptr`) [11] to manage resource lifecycles via RAII. While effective, C++ smart pointers are implemented as library-level wrappers rather than core language constructs. This library-level approach permits dangerous behaviors, such as implicit copies of un-encapsulated structures, silent null pointer dereferences, and the escape of raw pointers. Toka integrates lifecycle semantics directly into the core type system and syntax. By enforcing exclusive sigils and compile-time null safety checking, Toka is designed to prevent the silent escapes and null dereferences that continue to plague modern C++ software.

## Generational Safety

Other systems languages, such as Vale [12], seek to achieve safety by combining static checks with generational reference tracking. While generational references provide safety with low overhead, they require minor runtime tracking and allocation metadata. Toka's PAL checker and resource morphology operate entirely at compile time, introducing zero runtime metadata or execution overhead.

## Explicit Allocation and Zig

Zig rejects implicit allocations and hidden control flows, requiring developers to pass allocators explicitly to functions. Toka shares Zig's philosophical commitment to physical transparency and explicit cost modeling. However, while Zig relies on manual memory management—relying on the programmer to prevent double-free and use-after-free errors—Toka integrates ownership and borrow checking rules into the core language compiler, mitigating these common errors at compile time without sacrificing explicit layout visibility.

## Linear Logic and Austral

Toka's ownership model and explicit move semantics (via `cede`) are deeply rooted in the theoretical foundations of linear logic [13] and its early application to software programming [14]. Modern languages like Austral implement strict linear type systems combined with capability-based security. Toka borrows the core invariant of linear types—specifically, that unique resources must be consumed exactly once—but integrates it into a pragmatic, C-like systems programming paradigm. This integration allows standard developers to leverage exclusive ownership tracking without adopting the strict monadic structures or pure linear constraints of more academic languages.

## Checked C and C-Family Extensions

Checked C introduces static and dynamic bounds checks to mitigate spatial memory safety issues in legacy C codebases. While Checked C successfully mitigates buffer overflows, it does not target temporal safety errors (such as use-after-free). Toka provides temporal safety out of the box through PAL and morphological unique pointers, while preserving low-level compatibility and the execution speed of native target code.

### Mainstream Ownership Evolution: Swift and Move-Only Types

Mainstream programming languages are increasingly adopting ownership and move-only types to mitigate safety and concurrency challenges. For example, Swift has recently introduced non-copyable types and explicit ownership annotations (e.g., `consuming`, `borrowing`) to prevent compiler-injected reference-counting overhead in performance-critical sections [15]. Toka aligns with this modern mainstream evolution, but elevates explicit ownership to a primary language feature. By enforcing these rules uniformly across all types and composite shapes, Toka avoids the syntactic complexity of having dual copyable and non-copyable paradigms in the same language.

## Limitations and Future Work

While the design of Toka's explicit resource semantics offers a pragmatic approach to native systems programming, the current implementation and checking infrastructure possess several boundaries.

### Compile-time Checking Boundaries

The current Provenance, Alias, and Lifetime (PAL) checker is primarily intra-procedural. While dependency routing successfully tracks returned references, highly complex, cross-closure borrow relationships or extremely dynamic, nested alias networks are not fully analyzed. In such scenarios, the compiler adopts a conservative posture, rejecting complex alias graphs at compile time to guarantee safety. Refining these static checks into a more expressive, path-sensitive analysis is a primary focus of our ongoing work.

### Compiler Self-Hosting

The current production compiler, `tokac`, is constructed in C++ utilizing LLVM libraries. Although we have developed a parallel compiler implementation written in Toka itself, the self-hosting transition is not yet complete. Currently, compiling the Toka-authored compiler requires the C++ compiler as the primary bootstrap tool. Achieving full self-hosting and ensuring that the Toka-in-Toka compiler can reliably compile its own source remains a critical engineering milestone.

### Ecosystem and Industrial Validation

As a nascent systems programming language, Toka lacks a comprehensive ecosystem of libraries, database connectors, and platform-specific bindings. Additionally, the language has not yet been validated in large-scale industrial projects (exceeding 100,000 lines of code). Such validation is essential to verify that the cognitive benefits of the Handle-Soul model scale effectively across large development teams.

## Conclusion

In this paper, we presented Toka, a native systems programming language that introduces *Explicit Resource Semantics* to resolve the tension between physical transparency and memory safety. At the core of Toka's design is the *Handle-Soul Duality*, which syntactically splits pointer identities (Handles) from their underlying value payloads (Souls). By using naked names for value manipulation and explicit prefix sigils for pointer operations, Toka aligns language syntax with daily systems programming needs.

Our design details a comprehensive resource vocabulary, comprising unique, shared, borrowed, and raw pointers, complemented by explicit mutation and nullability markers. Through a pragmatic static checker (PAL) and composite shape boundaries, Toka is designed to reject typical classes of lifecycle and borrow violations at compile time without introducing complex lifetime parameter annotations. Our preliminary evaluation suggests that the proposed model can statically reject representative classes of spatial and temporal memory errors while preserving the zero-cost abstraction philosophy, presenting a viable alternative for safe, low-level systems programming.

## Acknowledgements

The author acknowledges the use of AI-assisted programming and writing tools during the development of the Toka compiler and the preparation of this manuscript. These tools were used for language refinement, technical discussion, and implementation assistance. All technical concepts, language designs, analyses, and conclusions presented in this work remain the sole responsibility of the author.

# Bibliography

[1] N. D. Matsakis and F. S. Klock II, “The Rust Language,” in *Proceedings of the 2014 ACM SIGAda Annual Conference on High Integrity Language Technology (HILT)*, 2014, pp. 103–104.

[2] A. M. Turing, “Checking a Large Routine,” in *Report of a Conference on High Speed Automatic Calculating Machines*, 1949, pp. 67–69.

[3] A. Weiss, D. Patterson, N. D. Matsakis, and A. Ahmed, “Oxide: The Essence of Rust,” *arXiv preprint arXiv:1903.00982*, 2019.

[4] R. Jung, J.-H. Jourdan, R. Krebbers, and D. Dreyer, “RustBelt: Securing the Foundations of the Rust Programming Language,” *Proceedings of the ACM on Programming Languages (PACMPL)*, vol. 2, no. POPL, pp. 66:1–66:34, 2018.

[5] D. Grossman, G. Morrisett, T. Jim, M. Hicks, Y. Wang, and J. Cheney, “Region-Based Memory Management in Cyclone,” in *Proceedings of the ACM SIGPLAN Conference on Programming Language Design and Implementation (PLDI)*, 2002, pp. 282–293.

[6] N. D. Matsakis, “Non-Lexical Lifetimes (NLL) in Rust.” 2018.

[7] The Rust Project, “Polonius: A Datalog-based Borrow Checker for Rust.” 2020.

[8] B. Liskov, A. Snyder, R. Atkinson, and C. Schaffert, “Abstraction Mechanisms in CLU,” *Communications of the ACM*, vol. 20, no. 8, pp. 564–576, 1977.

[9] D. G. Clarke, J. M. Potter, and J. Noble, “Ownership Types for Flexible Alias Protection,” in *Proceedings of the 13th ACM SIGPLAN Conference on Object-Oriented Programming, Systems, Languages, and Applications (OOPSLA)*, 1998, pp. 48–64.

[10] D. Racordon, D. Shabalin, D. Zheng, D. Abrahams, and B. Saeta, “Implementation Strategies for Mutable Value Semantics,” *Journal of Object Technology*, vol. 21, no. 2, pp. 1–15, 2022, doi: 10.5381/jot.2022.21.2.a2.

[11] S. Meyers, *Effective Modern C++: 42 Specific Ways to Improve Your Use of C++11 and C++14*. O'Reilly Media, Inc., 2014.

[12] E. Ovadia, “Generational References: Safe, Zero-Cost Memory Management in Vale.” 2020.

[13] J.-Y. Girard, “Linear Logic,” *Theoretical Computer Science*, vol. 50, no. 1, pp. 1–101, 1987.

[14] H. G. Baker, “Lively Linear Lisp — Look Ma, No Garbage!,” *ACM SIGPLAN Notices*, vol. 27, no. 8, pp. 89–98, 1992.

[15] J. Groff, “Roadmap for Improving Swift Performance Predictability: Ownership and Move-Only Types.” 2022.